\documentclass[seceq,preprint]{ptptex}

\usepackage{wrapft}
\usepackage{graphicx,color}

\title{Skyrme Strings}
\author{Muneto \textsc{Nitta}$^{1,}$\footnote{E-mail: nitta@phys-h.keio.ac.jp} 
and Noriko \textsc{Shiiki}$^{2,}$\footnote{E-mail: norikoshiiki@mail.goo.ne.jp}
}

\inst{$^1$Department of Physics, Keio University, Hiyoshi, Yokohama 
223-8521, Japan\\
$^2$Department of Management, Atomi University, Niiza 
352-8501, Japan}

\recdate{September 19, 2007}

\abst{We construct nontopological string solutions with $U(1)$ Noether charge 
in the Skyrme model with a pion mass term, and examine their stability 
by taking linear perturbations.
The solution exhibits a critical 
angular velocity beyond which the configuration energetically prefers to decay 
by emitting pions. This critical point is observed as a cusp in the relation 
between energy and charge. 
We find that the maximum length for the string to be stable is comparable 
to the size of one skyrmion. Beyond the length, it is unstable to decay. 
This instability raises the possiblity of dynamical realization of 
Skyrme strings from monopole strings inside a domain wall.
}
\begin{document}

\maketitle 

%%%%%%%%%%%%%%%%%%%%%%%%%%%%%%%%%%%%%
\section{Introduction}
The detection of the signal of chiral symmetry breaking in the 
early Universe or corresponding high-energy experiments is one of 
the most intriguing topics of research aimed at understanding quantum chromo 
dynamics (QCD). 
The spontaneous breaking of the chiral symmetry accompanies the production 
of massless Nambu-Goldstone bosons known as mesons. 
At low energies, QCD can be reduced to an effective theory described 
by only the meson degrees of freedom, called the sigma model.   
 
The linear sigma model contains heavy sigma particles as well as mesons 
so as to respect $SU(2)\times SU(2) \sim O(4)$ symmetry~\cite{Schwinger:1957em}. 
In the low-energy limit, one can integrate out the heavy sigma particles, and 
the nonlinear sigma model described by only pion fields is obtained. 
In this model, the sigma particles are dynamically generated as a bound state of two 
pions. Taking into account the terms up to the 4th-order derivative (Skyrme term) 
in the nonlinear sigma model, one can obtain the Skyrme model where topological 
soliton solutions, called skyrmions, are interpreted as baryons, and hadrons 
are described in a unified manner~\cite{Skyrme:1961vq}.  

A generic phenomenon expected as a consequence of the chiral phase 
transition is the formation of topological or nontopological defects via the Kibble 
mechanism~\cite{Balachandran:2001qn,Balachandran:2002je}.  
In fact, it has been shown that the linear sigma model yields nontopological 
string solutions composed of neutral pions and sigma particles
~\cite{Zhang:1997is,Brandenberger:1998ew,Mao:2004ym}.  
They are not topologically stable and hence decay into pions and sigma particles 
which subsequently decay into pions. The detailed study of the decay process will 
give insight into the chiral phase transition observed in, for example, heavy-ion 
collision experiments.  

The Skyrme model has been also known to possess nontopological string-like 
solutions~\cite{Jackson:1988xk,Jackson:1988ku}. 
The solutions are topologically unstable to decay. 
They are formed by the separation of the baryon density and may be closely 
related to QCD strings of quark-antiquark pairs. 
The possible decay modes of the string are many, and produce different numbers 
of mesons and baryon-antibaryon pairs. 
If pion strings in the linear sigma model are to be produced 
during the chiral phase transition, it is natural to expect 
that the Skyrme strings would also be produced in the low-energy regime. 
Although the Skyrme model describes baryons only approximately, we consider  
it to be a very convenient framework to incorporate baryons in the study 
of the chiral phase transition, and the string solutions would be worth further 
investigation.    

In this study, we extend the idea of the strings in the Skyrme model 
obtained by Jackson~\cite{Jackson:1988xk,Jackson:1988ku} to 
those with steady $U(1)$ rotation in the internal 
space~\cite{Coleman:1985ki,Ward:2003un}, obtain numerical Q-string 
solutions, and examine their stability. 
We find the critical length of the string is of the order 
of the effective length of one skyrmion. 
For the dynamical decay process of the Skyrme string, 
our solutions may be more interesting because the Q-string would decay into 
rotating baryon-antibaryon pairs, which are more realistic states than the 
static one~\cite{Rajaraman:1985ty,Battye:2005nx}.

Let us mention that string solutions in the baby-Skyrme model were previously 
obtained~\cite{Gisiger:1995yb} and discussed~\cite{Battye:1998zn}. 
The former indicates that the string solution is stable against decay into single baryons 
as it contains the energy per unit length less than the energy of an isolated baryon. 
The latter shows the possiblity of the reconnection of the strings.

\section{String solutions in Skyrme Model}
The Skyrme Lagrangian with the pion mass is defined by 
\begin{eqnarray}
	{\cal L}=\frac{F_{\pi}^{2}}{4}\,{\rm tr}\,(R_{\mu}R^{\mu})
	+\frac{1}{32e^{2}}\,{\rm tr}\,[R_{\mu},R_{\nu}]^{2}
	+\frac{1}{2}m_{\pi}^{2}F_{\pi}^{2}\,{\rm tr}\,(U-1),
	\label{skyrme_lag}
\end{eqnarray}
where $R_{\mu}=U^{\dagger}\partial_{\mu}U$ and $U$ is an $SU(2)$-valued chiral 
field given by
\begin{eqnarray}
	U=\phi_{0}+i {\vec \phi}\cdot {\vec \tau} \;\;\; {\rm with}
	\;\;\; \phi_{0}^{2}+{\vec \phi}^{2}=1 
\end{eqnarray}
and $F_{\pi}\sim 93$MeV is the pion decay constant, 
$m_{\pi}$ is the pion mass and $e$ is a free parameter whose value  
is about $5.45$ as given in Ref.~\cite{Adkins:1983ya} , for example. 

When $m_{\pi}=0$, Lagrangian~(\ref{skyrme_lag}) is invariant 
under the chiral symmetry $SU(2)_{L}\times SU(2)_{R}$ defined by $U \to U'=g_{L}Ug_{R}^{\dagger}$ 
with $g_{L}\in SU(2)_{L}$ and $g_{R}\in SU(2)_{R}$. The pion mass term 
explicitly breaks $SU(2)_{L}\times SU(2)_{R} \to SU(2)_{V}$ 
with $g_{L}=g_{R}$. 
We consider the $U(1)$ subgroup of $SU(2)_{V}$ with the transformation
\begin{eqnarray*}
	&& \phi_{1}\rightarrow \phi_{1}\cos \alpha -\phi_{2}\sin\alpha \,, \\
	&& \phi_{2}\rightarrow \phi_{1}\sin \alpha +\phi_{2}\cos\alpha \,.
	\label{u1_rot}	
\end{eqnarray*}
The associated U(1) current is 
\begin{eqnarray}
	J_{\mu}=\frac{\partial \delta{\cal L}}{\partial (\partial^{\mu}\alpha)}
	=\frac{F_{\pi}^{2}}{2}{\rm tr}(R_{\mu}A)
	+\frac{1}{8e^{2}}{\rm tr}([R_{\mu},R_{\nu}][A,R_{\nu}]), 
\end{eqnarray}
where 
\begin{eqnarray*}
	A=\left(
	\begin{array}{cc}
	i(\phi_{1}^{2}+\phi_{2}^{2}) & (\phi_{0}-i\phi_{3})(\phi_{1}-i\phi_{2}) \\
	-(\phi_{0}+i\phi_{3})(\phi_{1}+i\phi_{2}) & -i(\phi_{1}^{2}+\phi_{2}^{2}) 
	\end{array}\right) \,.
\end{eqnarray*}
The conserved U(1) charge per unit length in the z-direction is given 
by the spatial integral of the zeroth component of the current,
\begin{eqnarray}
	Q=\int dx\,dy \, J^{0} \,. \label{charge}
\end{eqnarray}
To obtain string solutions with the $U(1)$ charge, let us consider 
the ansatz constructed by Jackson~\cite{Jackson:1988xk,Jackson:1988ku} 
and induce steady rotation in the internal space by setting  
\begin{eqnarray}
	\alpha =\alpha (t)
\end{eqnarray}
in the $U(1)$ transformation of Eq.~(\ref{u1_rot}). Then we have    
\begin{eqnarray}
	U=\left(
	\begin{array}{cc}
	\cos f(r) & i\sin f(r)e^{-i(\theta+\alpha(t))} \\
	i\sin f(r) e^{i(\theta+\alpha(t))} & \cos f(r) 
	\end{array}
	\right)
	\label{ansatz}
\end{eqnarray}
in the cylindrical coordinate system with the metric 
\begin{eqnarray}
	ds^{2}=-dt^{2}+dz^{2}+dr^{2}+r^{2}d\theta^{2},
\end{eqnarray}
where ${\hat r}^{i}$ is a unit vector in the direction of $r$. 
This ansatz associates rotation in isospace with rotation in space. 
 
Substituting ansatz~(\ref{ansatz}) into Eq.~(\ref{charge}), one obtains 
\begin{eqnarray}
	Q= \frac{2\pi F_{\pi}}{e}\int d\rho \,\rho \sin^{2}f (1+f'^{2}) 
	\,{\dot \alpha},
\end{eqnarray}
where we have rescaled $\rho \equiv e F_{\pi}r$ and $\tau \equiv 
eF_{\pi}t$, and the prime and the dot denote differentiation with respect 
to $\rho$ and $\tau$, respectively.
 
To find the minimum of the string tension for fixed $Q$, 
we introduce a Lagrange 
multiplier $\omega$ and write the string tension in terms of $Q$ as~\cite{Kusenko:1997ad}  
\begin{eqnarray}
	{\cal E}_{\omega}&=& {\cal T}+{\cal V}+{\hat \omega}[{\hat Q}
	-2\pi\int d\rho \,\rho \sin^{2}f(1+f'^{2})\,{\dot \alpha}] \nonumber \\
	&=& {\cal V}+\pi\int d\rho \,\rho \sin^{2}f(1+f'^{2})
	({\dot \alpha}^{2}-2{\hat \omega}{\dot \alpha}) +{\hat \omega}{\hat Q} \nonumber \\
	&=& {\cal V}-\pi{\hat \omega}^{2}\int d\rho \, \rho\sin^{2}f(1+f'^{2}) 
	+\pi\int d\rho\,\rho\sin^{2}f (1+f'^{2})
	({\dot \alpha}-{\hat \omega})^{2} \nonumber \\
	&& + {\hat \omega}{\hat Q}\label{energy}
\end{eqnarray}
where we have defined 
\begin{eqnarray}
	{\cal T}&\equiv&\pi \int d\rho\,\rho\sin^{2}f (1+f'^{2})\,
	{\dot \alpha}^{2}, \\
	{\cal V}&\equiv&\pi \int d\rho \,\rho\left[\left(1+\frac{\sin^{2}f}{\rho^{2}}
	\right)f'^{2}+\frac{\sin^{2}f}{\rho^{2}}+2{\hat m}_{\pi}^{2}
	(1-\cos f)\right] 
\end{eqnarray}
with ${\hat m}_{\pi}\equiv m_{\pi}/eF_{\pi}$, 
${\hat \omega}\equiv \omega /eF_{\pi}$, 
${\hat Q}\equiv eQ/F_{\pi}$ and 
${\cal E}_{\omega}\equiv eE/F_{\pi}$. 
The third term in Eq.~(\ref{energy}), which is the only time-dependent term,  
is positive definite and therefore should vanish at the minimum. 
We thus set $\alpha = {\hat \omega}\tau$ and the tension can be 
written as    
\begin{eqnarray}
	{\cal E}_{\omega}={\cal V}+\frac{{\hat Q}^{2}}{2{\cal I}} \label{classical_en}
\end{eqnarray}
with the moment of inertia for iso-rotation being
\begin{eqnarray*}
	{\cal I}=2\pi \int d\rho\, \rho\sin^{2}f (1+f'^{2})\,.
\end{eqnarray*}
One can see that for fixed ${\hat Q}$, the charge term ${\hat Q}^{2}/2{\cal I}$ 
plays the role of stabilizing string solutions as it is inversely proportional 
to the functional of the profile $f$.
  
The field equation can be obtained by taking the variations 
of $f$ in string tension~(\ref{energy}),  
\begin{eqnarray}
	&& \left(1-{\hat \omega}^{2}\sin^{2}f+\frac{\sin^{2}f}{\rho^{2}}\right)f''
	+\left(1-{\hat \omega}^{2}\sin^{2}f-\frac{\sin^{2}f}{\rho^{2}}\right) 
	\frac{f'}{\rho} \nonumber \\ 
	&& +\frac{\sin f\cos f}{\rho^{2}}(1-{\hat \omega}^{2}\rho^{2})
	(f'^{2}-1) -{\hat m}_{\pi}^{2}\sin f =0. \label{skyrme_eq}
\end{eqnarray}
Note that to find full solutions in our system,  
we must solve the equation of motion without imposing 
any ansatz. In this sense, the solution of Eq.~(\ref{skyrme_eq}) 
may not be the true minimum of the action. It, however, minimises 
the action within the cylindrical symmetric configuration.
The finiteness and regularity of string tension require the boundary conditions   
\begin{eqnarray}
	f(\infty) = 0, \;\;\; f(0)=n\pi, \label{boundary}
\end{eqnarray}
where $n$ is any integer. In this paper, we only examine the case 
of $n=1$. Other values of $n$ will be reported elsewhere. 
Equation.~(\ref{skyrme_eq}) is solved numerically subject to boundary 
conditions~(\ref{boundary}). 
  
The asymptotic form of the profile $f(\rho)$ as $\rho \rightarrow \infty$ can be 
obtained by linearizing the field equation. Setting $f=\delta f$, one can get  
\begin{eqnarray}
	\delta f''+\frac{1}{\rho}\delta f'-\frac{1}{\rho^{2}}\delta f 
	-\left({\hat m}_{\pi}^{2}-{\hat \omega}^{2}\right)
	\delta f=0\, \label{linear}
\end{eqnarray}
with the solution of the Bessel function 
\begin{eqnarray*}
	\delta f = CK_{1}(b\rho), 
\end{eqnarray*}
where $C$ is an arbitrary constant and 
\begin{eqnarray}
	b=\sqrt{{\hat m}_{\pi}^{2}-{\hat \omega}^{2}} \,.
\end{eqnarray}
This restricts the value of ${\hat \omega}$ as   
\begin{eqnarray}
	0<{\hat \omega}<{\hat m}_{\pi}\,.
\end{eqnarray}
Thus, there exists a critical value ${\hat \omega}_{+}={\hat m}_{\pi}$ 
beyond which soliton solutions cannot be found. This is because for ${\hat \omega}
={\hat \omega}_{+}$, $f \sim 1/\rho$ and for ${\hat \omega}
>{\hat \omega}_{+}$, $f\sim CJ_{1}(b'\rho)$ with $b'=\sqrt{{\hat \omega}^{2}
-{\hat \omega}_{+}^{2}}$ as $\rho \rightarrow \infty$, resulting in the divergent 
string tension and inertia moment $I$.   
Physically, this oscillatory behaviour of $f$ signals the instability of 
the string against the emission of pions.
 
String solutions are obtained by solving Eq.~(\ref{skyrme_eq}) numerically.  
Figure~\ref{fig:f} shows the profile $f$ as a function of $\rho$ for several values 
of ${\hat \omega}$. As the value of ${\hat \omega}$ increases, the size of the soliton 
expands. This expansion is due to the centrifugal force effect. 
Figure~\ref{fig:q-e} shows the string tension as a function of the charge. 
The tension increases as the charge and/or the pion mass increases. 
The approximate asymptotic formula of the tension can be deduced analytically 
as~\cite{Piette:1994mh}  
\begin{eqnarray}
	{\cal E}_{\omega}\sim {\rm const}+\frac{{\hat m}_{\pi}}{2}\,{\hat Q}\,, \label{}
\end{eqnarray}
which holds in our numerical results within a few percent error. 
We found the cusp appears at the critical value of ${\hat \omega}$.  
In the context of Q-ball solutions, the second branch represents unstable solutions 
called Q-clouds~\cite{Alford:1987vs}. 
Consistently, our second branch represents string solutions which energetically 
favour decay by the emission of pions.

\section{Linear stability analysis}
In this section, we shall examine the linear stability of our string solutions 
obtained in the previous section to see whether the $U(1)$ rotation changes their 
stability.   
To study the linear stability of the soliton solution, let us consider an 
infinitesimal fluctuation in the $x_{3}$-direction in the internal space: 
\begin{eqnarray}
	U=\left(
	\begin{array}{cc}
	\cos f + i\delta_{3} & ie^{-i(\theta+{\hat \omega}\tau)}\sin f \\
	ie^{i(\theta+{\hat \omega} \tau)}\sin f & \cos f - i\delta_{3} 
	\end{array}
	\right) , \label{time_u}
\end{eqnarray}
where $\delta_{3}$ is the fluctuation. 
The field $U$ in Eq.~(\ref{time_u}) is then unitary up to the first order 
in $\delta_{3}$.
Note that one can show the fluctuations in other internal directions 
decouple from $\delta_{3}$ and contribute only to raising the total 
energy of the configuration~\cite{Jackson:1988ku}. 

The field equation for $U$ is derived as 
\begin{eqnarray}
	\partial_{\mu}R_{\mu}^{i}+\partial_{\mu}[R_{\nu}^{j}
	(R_{\mu}^{i}R_{\nu}^{j}-R_{\mu}^{j}R_{\nu}^{i})]
	+\frac{1}{2}{\hat m}_{\pi}^{2}{\rm tr}(i\tau^{i}U)=0,
	\label{g-eq}
\end{eqnarray}
where we have defined $U^{\dagger}\partial_{\mu}U=iR_{\mu}^{i}\tau^{i}$.
Inserting Eq.~(\ref{time_u}) into Eq.~(\ref{g-eq}) and taking the first-order 
terms in $\delta_{3}$, one obtains the following equation for $\delta_{3}$:  
\small{
\begin{eqnarray}
	&& \left(1-{\hat \omega}^{2}\sin^{2}f+\frac{\sin^{2}f}{\rho^{2}}\right)
	\delta_{3}''+\left[-2{\hat \omega}^{2}f'\sin f\cos f+\frac{1}{\rho}
	\left\{1-{\hat \omega}^{2}\sin^{2}f +\frac{\sin f}{\rho}\left(
	2f'\cos f-\frac{\sin f}{\rho}\right)\right\}\right]\delta_{3}'\nonumber \\
	&& +\frac{1}{\rho^{2}}(1+f'^{2}-{\hat \omega}^{2}\sin^{2}f)
	\partial_{\theta}^{2}\delta_{3}
	+\left(1+f'^{2}-{\hat \omega}^{2}\sin^{2}f+\frac{\sin^{2}f}{\rho^{2}}
	\right)\partial_{z}^{2}\delta_{3}+2{\hat \omega}
	\frac{\sin^{2}f}{\rho^{2}}\partial_{\theta}{\dot \delta}_{3} \nonumber \\
	&& +\left[f'^{2}-{\hat \omega}^{2}(1+2f'^{2})\sin^{2}f+(1+2f'^{2})
	\frac{\sin^{2}f}{\rho^{2}}-{\hat m}_{\pi}^{2}\cos f \right]\delta_{3} 
	=\left[1+(1+{\hat \omega}^{2})\left(f'^{2}+\frac{\sin^{2}f}{\rho^{2}}\right)
	\right]{\ddot \delta}_{3}. \label{eq_del}
\end{eqnarray}}
Let us define the length of the string in the $z$ direction as $L$. Then 
the boundary conditions at $z=\pm \frac{1}{2}L$ are given by 
\begin{eqnarray}
	\delta_{3}(\rho ,\theta, z=\pm L/2) = 0\,.
\end{eqnarray}
Setting 
\begin{eqnarray}
	\delta_{3}=e^{i\Omega \tau}e^{im\theta}\cos (k_{z}z)R (\rho),
\end{eqnarray}
where $m$ is an integer and 
\begin{eqnarray}
	k_{z}=\frac{(2n+1)\pi}{L},  \label{kz} 
\end{eqnarray}
with $n$ being an integer, and considering the most unstable mode $m=n=0$, 
Eq.~(\ref{eq_del}) is reduced to 
\small{ 
\begin{eqnarray}
	&& -\left(1-{\hat \omega}^{2}\sin^{2}f+\frac{\sin^{2}f}{\rho^{2}}\right)
	R''-\left[-2{\hat \omega}^{2}f'\sin f\cos f+\frac{1}{\rho}\left\{
	1-{\hat \omega}^{2}\sin^{2}f+\frac{\sin f}{\rho}\left(2f'\cos f
	-\frac{\sin f}{\rho}\right)\right\}\right]R' \nonumber \\
	&& +\left[\left(1+f'^{2}-{\hat \omega}^{2}\sin^{2}f+\frac{\sin^{2}f}{\rho^{2}}
	\right) \left(\frac{\pi}{L}\right)^{2}-\left(1-2{\hat \omega}^{2}\sin^{2}f
	+\frac{2\sin^{2}f}{\rho^{2}}\right)f'^{2}-\frac{\sin^{2}f}{\rho^{2}}
	(1-{\hat \omega}^{2}\rho^{2})+{\hat m}_{\pi}^{2}\cos f \right]R \nonumber \\
	&& =\left(1+f'^{2}+\frac{\sin^{2}f}{\rho^{2}}\right)\Omega^{2}R \,. 
	\label{eq_r}
\end{eqnarray}}
We introduce a new coordinate ${\tilde \rho}$ such that 
\begin{eqnarray}
	\frac{d{\tilde \rho}}{d\rho}= \left[\rho\left(1-{\hat \omega}^{2}
	\sin^{2}f+\frac{\sin^{2}f}{\rho^{2}}\right)\right]^{-1}. \label{}
\end{eqnarray}
Then Eq.~(\ref{eq_r}) takes the form of the Sturm-Liuville equation   
\begin{eqnarray}
	-\frac{d^{2}R}{d{\tilde \rho}^{2}}+V R =
	\rho \left(1+f'^{2}+\frac{\sin^{2}f}{\rho^{2}}
	\right)\Omega^{2}R \,, \label{}
\end{eqnarray}
where 
\begin{eqnarray}
	V&=&\rho \left[\left(1+f'^{2}-{\hat \omega}^{2}\sin^{2}f
	+\frac{\sin^{2}f}{\rho^{2}}\right) \left(\frac{\pi}{L}\right)^{2}
	+\left(1-2{\hat \omega}^{2}\sin^{2}f+\frac{2\sin^{2}f}{\rho^{2}}
	\right)f'^{2} \right. \nonumber \\
	&& \left. +\frac{\sin^{2}f}{\rho^{2}}(1-{\hat \omega}^{2}\rho^{2})
	-{\hat m}_{\pi}^{2}\cos f \right] . \label{}
\end{eqnarray}
The solution is linearly stable if there are no normalizable modes with 
negative energy (bound states), because such modes realise exponentially diverging 
$\delta_{3}$ owing to imaginary $\Omega$~\cite{Luckock:1986tr,Heusler:1991xx,
Bizon:1992gb,Gibbons:2002pq}. 

It is straightforward to show that the normalizable wave function should take 
an asymptotic form of the Bessel function for large $\rho$, 
\begin{eqnarray}
	R(\rho)\sim K_{0}(\kappa_{1}\rho) \;\;\;\;\; {\rm with} \;\;\; 
	\kappa_{1}=\sqrt{\left(\frac{\pi}{L}\right)^{2}+{\hat m}_{\pi}^{2}
	-\Omega^{2}},
\end{eqnarray}
and for small $\rho$,
\begin{eqnarray}
	R(\rho)\sim J_{0}(\kappa_{2}\rho) \;\;\;\;\; {\rm with} \;\;\; 
	\kappa_{2}=\sqrt{\frac{(1+2f_{1}^{2})(\Omega^{2}-\pi^{2}/L^{2})
	+{\hat m}_{\pi}^{2}+2(1+f_{1}^{2})f_{1}^{2}}{1+f_{1}^{2}}},
\end{eqnarray}
where we have defined $f_{1}=f'(0)$. 

In order for $\xi$ to be normalizable, $\kappa_{1}$ and  
$\kappa_{2}$ must be real, which gives an inequality for $\Omega^{2}$:    
\begin{eqnarray}
	\left(\frac{\pi}{L}\right)^{2}-\frac{1}{1+2f_{1}^{2}}[{\hat m}_{\pi}^{2}
	+2(1+f_{1}^{2})f_{1}^{2}]\, <\Omega^{2}<\, \left(\frac{\pi}{L}\right)^{2}
	+{\hat m}_{\pi}^{2}. \label{normalisable}
\end{eqnarray}
The condition that there exists no tachionic mode is $\Omega^{2}>0$, that is, 
the left-hand side in Eq.~(\ref{normalisable}) is positive,  
\begin{eqnarray}
	\left(\frac{\pi}{L}\right)^{2}-\frac{1}{1+2f_{1}^{2}}[{\hat m}_{\pi}^{2}
	+2(1+f_{1}^{2})f_{1}^{2}] >0 \,.  \label{}
\end{eqnarray}
This leads to the constraint for the length of the stable string: 
\begin{eqnarray}
	L \, < \, \sqrt{\frac{(1+2f_{1}^{2})\pi^{2}}{{\hat m}_{\pi}^{2}+2(1+f_{1}^{2})
	f_{1}^{2}}} \,. 
\end{eqnarray}
In evaluating $f_{1}$ numerically, we have $f_{1} \sim O(1)$. 
Roughly speaking, ${\hat m}_{\pi}=m_{\pi}/eF_{\pi} =137/(5.45\times 93)=0.27  
 \sim O(10^{-1})$ and hence, $L \sim O(1)$.  
As the mass of one skyrmion $M_{s}/eF_{\pi}= 1000/(5.45\times 93) 
= 1.97 \sim O(1)$, the maximum length for the string to 
be stable is of the order of the length of one skyrmion. 
When the string becomes longer than that, it will decay by emitting pions. 
%One may recall the fact that in nuclear physics, the Compton wavelength of 
%the pion is of the same order as the size of a Skyrmion. 
We have also checked this conclusion by solving Eq.~(\ref{eq_r}) 
numerically for both vanishing and nonvanishing values of ${\hat \omega}$.    

%%%%%%%%%%%%%%%%%%%%%%%%%%%%%%%%%%
\section{Conclusions}
In this paper we constructed string solutions with the $U(1)$ Noether charge 
in the Skyrme model and examined whether the $U(1)$ charge 
could stabilize the string solution. 

\begin{figure}
\begin{center}
\includegraphics[height=6.5cm, width=8.0cm]{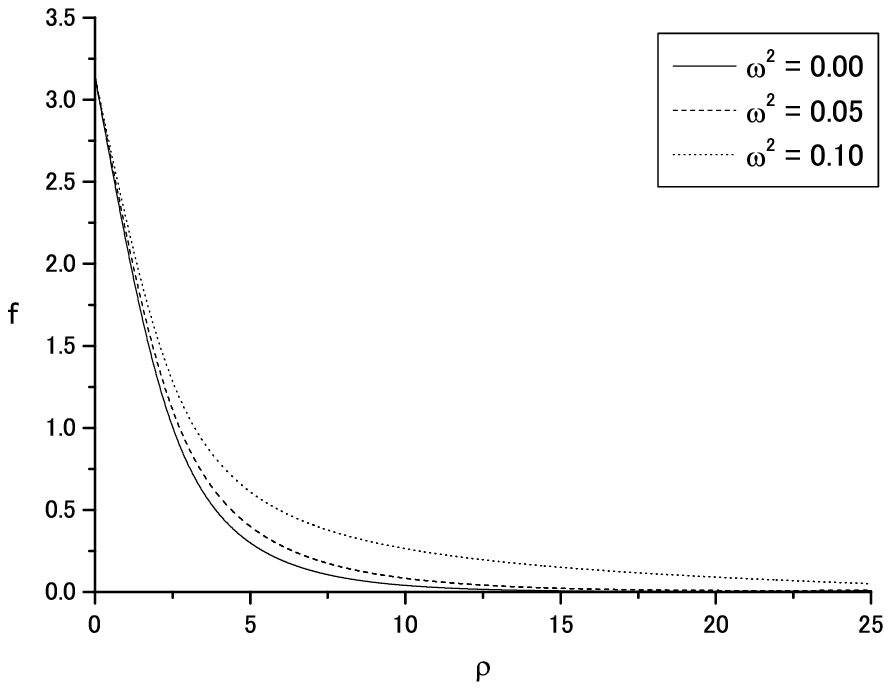}
\caption{\label{fig:f} Profile function $f$ as a function of $\rho$ with 
${\hat m}_{\pi}^{2}=0.1$ for ${\hat \omega}=0.00,0.05,0.10$.}
\includegraphics[height=6.5cm, width=8.0cm]{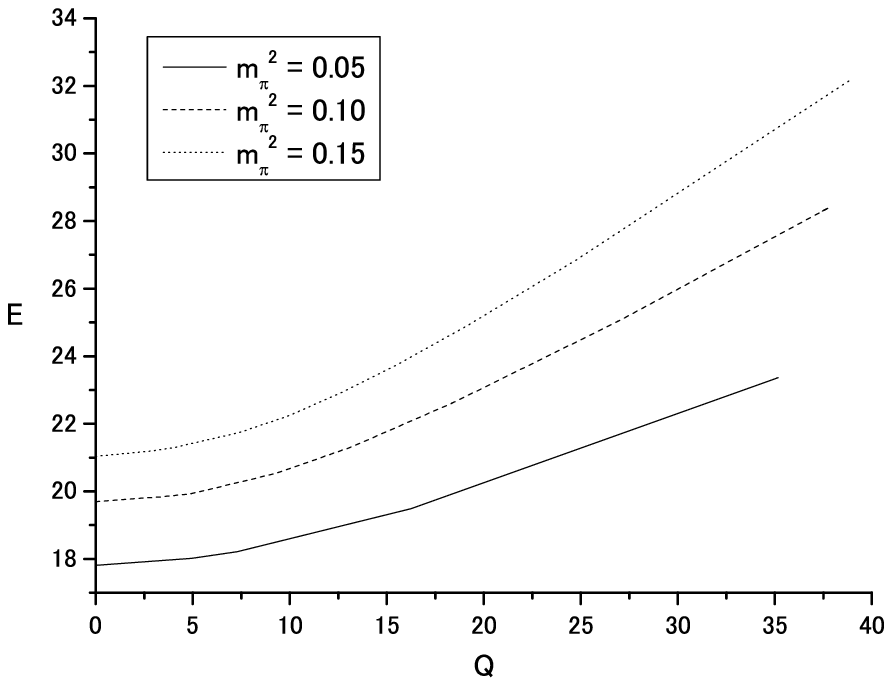}
\caption{\label{fig:q-e} Tension ${\cal E}_{\omega}$ of string  
as a function of charge ${\hat Q}$ for ${\hat m}_{\pi}^{2}=0.05,0.10,0.15$.}
\includegraphics[height=6.5cm, width=8.0cm]{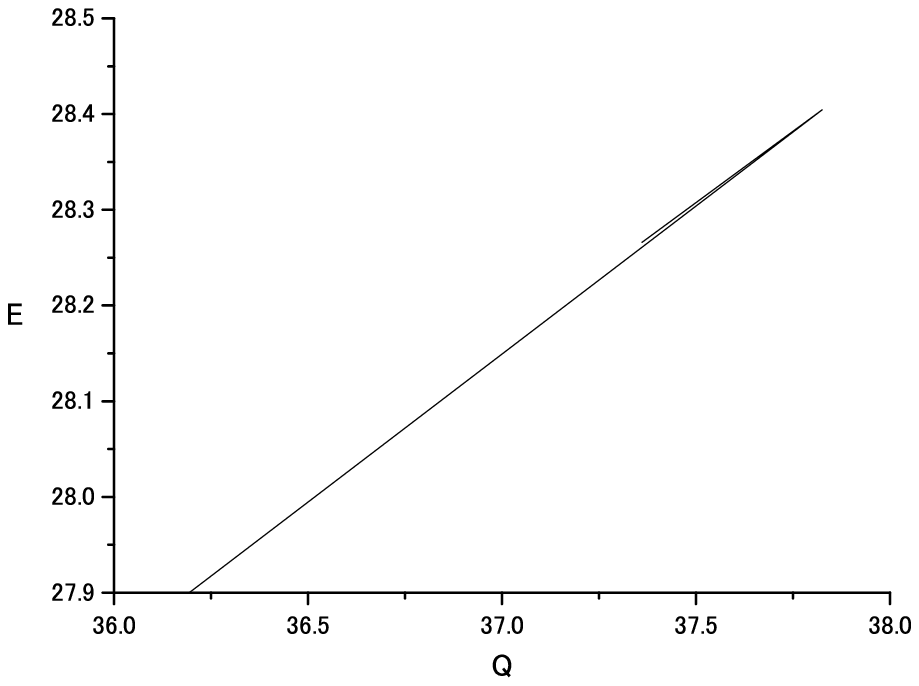}
\caption{\label{fig:cusp} Tension ${\cal E}_{\omega}$ of string  
as a function of charge ${\hat Q}$ for ${\hat m}_{\pi}^{2}=0.10$. 
The cusp is observed at the critical value of ${\hat \omega}$ which is 
${\hat \omega}_{+}={\hat m}_{\pi}$.}
\end{center}
\end{figure}

The string solution is exponentially localized in the radial direction if the 
angular frequency is less than the pion mass. Otherwise it is oscillatory along 
the radial direction.  
We found that there exists a critical value of the angular velocity beyond which 
the solution energetically favours decay by the emission of pions. 
This unstable branch of the string is observed as a cusp in the relation 
between the energy and the charge, as shown in Figure~\ref{fig:cusp}. 

The stability was examined by taking linear perturbations. 
We found that the maximum length for the string to be stable is comparable 
to the size of a skyrmion. Beyond that length, they are unstable to decay. 

A similar instability is observed in the case of skyrmions 
when one performs the projection after $SU(2)$-isorotation (projection 
after rotation)~\cite{Braaten:1984qe,Biedenharn:1984he,hajduk84,liu84,hayashi84}. 
Our string solutions have also been obtained by the projection after 
$U(1)$-isorotation. In this method, the configuration breaks down completely 
in the chiral limit ($m_{\pi}=0$) owing to the radiation of pion waves.
 
When the dynamical decay process of the Skyrme string is considered,  
our solutions may be more interesting because these would decay into 
rotating baryon-antibaryon pairs, which are more realistic states than the 
static one~\cite{Rajaraman:1985ty,Battye:2005nx}. 
In particular, if the Skyrme strings are assumed to be produced 
during the chiral phase transition, as are pion strings, the study of the decay 
mode will provide important information on high-energy experiments. 

Our solutions are considered to be the embedding of the (2+1)-dimensional 
spinning skyrmion in 3+1 dimensions and therefore, they are 
qualitatively similar to the solutions reported in~\cite{Piette:1994mh}.
When topological or nontopological solitons are embedded in higher dimensional 
spacetime, the instability in the direction of worldvolume is not alleviated   
by the existence of a Noether conserved charge.  
Thus, an alternative mechanism of stabilizing the Skyrme strings should be 
considered, such as cosmological expansion, which is a technique 
of stabilizing $\sigma$-model lump strings~\cite{Ward:2002ci}.
The Skyrme string might, however, be physically advantageous over 
the lump string in that the former is radially stable up to the effective length 
of one skyrmion while the latter is unstable to collapse.  
Thus it would be interesting to study them in the context of cosmic strings 
or to discuss the Skyrme strings in the context of the QCD strings since we are 
not yet sure how they are actually relevant to QCD strings.    

Approximate skyrmion solutions are known to be obtained from a holonomy of 
the Yang-Mills instanton particles in five space-time dimensions \cite{Atiyah:1989dq}. 
It has recently been shown that this situation can be realized by placing 
instanton particles inside a domain wall whose low-energy dynamics is described 
by the Skyrme model \cite{Eto:2005cc}. 
In the same way, unstable Skyrme strings discussed in this paper may be 
approximately constructed from a holonomy of monopole strings inside 
the domain wall, because it has been shown by Eto $et$ $al$. \cite{Eto:2005sw} 
that a monopole and a domain wall cannot coexist as a 
Bogomol'nyi-Prasad-Sommerfield state and the configuration is unstable. 

Finally, the possible realization of 
our solution in string theory, such as AdS/QCD \cite{Sakai:2004cn} 
where skyrmions 
are constructed \cite{Nawa:2006gv}, will be very interesting to explore.

%%%%%%%%%%%%%%%%%%%%%%%%%%%%%%%%%%


\newpage
\begin{thebibliography}{99}

\bibitem{Schwinger:1957em}
J.~S.~Schwinger, Ann. of Phys.\  {\bf 2} (1957), 407.

\bibitem{Skyrme:1961vq}
T.~H.~R.~Skyrme, Proc.\ Roy.\ Soc.\ London \  A {\bf 260} (1961), 127.

\bibitem{Balachandran:2001qn}
A.~P.~Balachandran and S.~Digal,
Int.\ J.\ Mod.\ Phys.\  A {\bf 17} (2002), 1149; hep-ph/0108086.

\bibitem{Balachandran:2002je}
A.~P.~Balachandran and S.~Digal, Phys.\ Rev.\  D {\bf 66} (2002), 
034018; hep-ph/0204262.

\bibitem{Zhang:1997is}
X.~Zhang, T.~Huang and R.~H.~Brandenberger,
Phys.\ Rev.\  D {\bf 58} (1998), 027702; hep-ph/9711452.

\bibitem{Brandenberger:1998ew}
R.~H.~Brandenberger and X.~M.~Zhang,  
Phys.\ Rev.\  D {\bf 59} (1999), 081301; hep-ph/9808306.

\bibitem{Mao:2004ym}
H.~Mao, Y.~Li, M.~Nagasawa, X.~M.~Zhang and T.~Huang, 
Phys.\ Rev.\  C {\bf 71} (2005), 014902; hep-ph/0404132.

\bibitem{Jackson:1988xk}
A.~Jackson, Nucl.\ Phys.\  A {\bf 493} (1989), 365.

\bibitem{Jackson:1988ku}
A.~Jackson, Nucl.\ Phys.\  A {\bf 496} (1989), 667.

\bibitem{Coleman:1985ki}
S.~R.~Coleman, Nucl.\ Phys.\  B {\bf 262} (1985), 263. 

\bibitem{Ward:2003un}
R.~S.~Ward, J.\ Math.\ Phys.\  {\bf 44} (2003), 3555; hep-th/0302045.

\bibitem{Rajaraman:1985ty}
R.~Rajaraman, H.~M.~Sommermann, J.~Wambach and H.~W.~Wyld, 
Phys.\ Rev.\  D {\bf 33} (1986), 287.

\bibitem{Battye:2005nx}
R.~A.~Battye, S.~Krusch and P.~M.~Sutcliffe, Phys.\ Lett.\  B 
{\bf 626} (2005), 120; hep-th/0507279.

\bibitem{Gisiger:1995yb}
T.~Gisiger and M.~B.~Paranjape, Phys.\ Lett.\  B {\bf 384} (1996), 207; 
hep-ph/9507223.

%\cite{Battye:1998zn}
\bibitem{Battye:1998zn}
R.~A.~Battye and P.~Sutcliffe, Proc.\ Roy.\ Soc.\ Lond.\  A 
{\bf 455} (1999), 4305; hep-th/9811077.

\bibitem{Adkins:1983ya}
G.~S.~Adkins, C.~R.~Nappi and E.~Witten, Nucl.\ Phys.\  B 
{\bf 228} (1983), 552.

\bibitem{Kusenko:1997ad}
A.~Kusenko, Phys.\ Lett.\  B {\bf 404} (1997), 285; hep-th/9704073.

\bibitem{Piette:1994mh}
B.~M.~A.~Piette, B.~J.~Schroers and W.~J.~Zakrzewski, 
Nucl.\ Phys.\  B {\bf 439} (1995), 205; hep-ph/9410256.

\bibitem{Alford:1987vs}
M.~G.~Alford, 
Nucl.\ Phys.\  B {\bf 298} (1988), 323.

\bibitem{Luckock:1986tr}
H.~Luckock and I.~Moss, Phys.\ Lett.\  B {\bf 176} (1986), 341.

\bibitem{Heusler:1991xx}
M.~Heusler, S.~Droz and N.~Straumann, Phys.\ Lett.\  B {\bf 271} (1991), 61; 
Phys.\ Lett.\  B {\bf 285} (1992), 21.

\bibitem{Bizon:1992gb}
P.~Bizon and T.~Chmaj, Phys.\ Lett.\  B {\bf 297} (1992), 55.

\bibitem{Gibbons:2002pq}
G.~Gibbons and S.~A.~Hartnoll, Phys.\ Rev.\  D {\bf 66} (2002), 064024; 
hep-th/0206202.

\bibitem{Braaten:1984qe}
E.~Braaten and J.~P.~Ralston, Phys.\ Rev.\  D {\bf 31} (1985), 598.

\bibitem{Biedenharn:1984he}
L.~C.~Biedenharn, Y.~Dothan and M.~Tarlini, Phys.\ Rev.\  D {\bf 31} 
(1985), 649.
 
\bibitem{hajduk84} 
C.~Hajduk and B.~Schwesinger, Phys.\ Lett.\  B {\bf 145} (1984), 171. 
 
\bibitem{liu84}
K.~F.~Liu and J.~S.~Zhang, Phys.\ Rev.\  D {\bf 30} (1984), 2015. 
 
\bibitem{hayashi84}
A.~Hayashi and G.~Holzwarth, Phys.\ Lett.\  B {\bf 140} (1984), 175.

\bibitem{Ward:2002ci}
R.~S.~Ward, Class.\ Quant.\ Grav.\  {\bf 19} (2002), L17; gr-qc/0201042.

%\cite{Atiyah:1989dq}
\bibitem{Atiyah:1989dq}
M.~F.~Atiyah and N.~S.~Manton, Phys.\ Lett.\  B {\bf 222} (1989), 438.

%\cite{Eto:2005cc}
\bibitem{Eto:2005cc}
M.~Eto, M.~Nitta, K.~Ohashi and D.~Tong, Phys.\ Rev.\ Lett.\  
{\bf 95} (2005), 252003; hep-th/0508130.

%\cite{Eto:2005sw}
\bibitem{Eto:2005sw}
M.~Eto, Y.~Isozumi, M.~Nitta and K.~Ohashi, 
Nucl.\ Phys.\  B {\bf 752} (2006), 140; hep-th/0506257.

%\cite{Sakai:2004cn}
\bibitem{Sakai:2004cn}
T.~Sakai and S.~Sugimoto, Prog.\ Theor.\ Phys.\  {\bf 113} (2005), 843; 
hep-th/0412141. \\
T.~Sakai and S.~Sugimoto, Prog.\ Theor.\ Phys.\  {\bf 114} (2006), 
1083; hep-th/0507073.

%\cite{Nawa:2006gv}
\bibitem{Nawa:2006gv}
K.~Nawa, H.~Suganuma and T.~Kojo, Phys.\ Rev.\  D {\bf 75} (2007), 086003; 
hep-th/0612187. \\
K.~Nawa, H.~Suganuma and T.~Kojo, Prog.\ Theor.\ Phys.\ Suppl.\ 
168 (2007), 231; hep-th/0701007. \\
H.~Hata, T.~Sakai, S.~Sugimoto and S.~Yamato, arXiv:hep-th/0701280.

%\cite{MacKenzie:2001av}
%\bibitem{MacKenzie:2001av}
%  R.~B.~MacKenzie and M.~B.~Paranjape,
%  ``From Q-walls to Q-balls,''
%  JHEP {\bf 0108}, 003 (2001)
%  [arXiv:hep-th/0104084].

%\cite{Leese:1991hr}
%\bibitem{Leese:1991hr}
%  R.~A.~Leese,
%  ``Q lumps and their interactions,''
%  Nucl.\ Phys.\  B {\bf 366}, 283 (1991); 
%  %%CITATION = NUPHA,B366,283;%%
%\cite{Abraham:1991ki}
%\bibitem{Abraham:1991ki}
%  E.~Abraham,
%  ``Nonlinear Sigma Models And Their Q Lump Solutions,''
%  Phys.\ Lett.\  B {\bf 278}, 291 (1992); 
%  %%CITATION = PHLTA,B278,291;%%
%\cite{Naganuma:2001pu}
%\bibitem{Naganuma:2001pu}
%  M.~Naganuma, M.~Nitta and N.~Sakai,
%  ``BPS lumps and their intersections in N = 2 SUSY nonlinear sigma models,''
%  Grav.\ Cosmol.\  {\bf 8}, 129 (2002)
%  [arXiv:hep-th/0108133];
  %%CITATION = GRCOF,8,129;%%
%\cite{Bak:2006qk}
%\bibitem{Bak:2006qk}
%  D.~Bak, S.~O.~Hahn, J.~Lee and P.~Oh,
%  ``Supersymmetric Q-lumps in the Grassmannian nonlinear sigma models,''
%  Phys.\ Rev.\  D {\bf 75}, 025004 (2007)
%  [arXiv:hep-th/0610067].
  %%CITATION = PHRVA,D75,025004;%%


\end{thebibliography}
\end{document}